# Acoustic attenuation in magnetic insulator films: effects of magnon polaron formation

Shihao Zhuang and Jia-Mian Hu*

*Department of Materials Science and Engineering, University of Wisconsin-Madison, Madison, WI, 53706, USA*

## Abstract

A magnon and a phonon are the quanta of spin wave and lattice wave, respectively, and they can hybridize into a magnon polaron when their frequencies and wavenumbers match close enough the values at the exceptional point. Guided by an analytically calculated magnon polaron dispersion, dynamical phase-field simulations are performed to investigate the effects of magnon polaron formation on the attenuation of a bulk acoustic wave in a magnetic insulator film. It is shown that a stronger magnon-phonon coupling leads to a larger attenuation. The simulations also demonstrate the existence of a minimum magnon-phonon interaction time required for the magnon polaron formation, which is found to decrease with the magnetoelastic coupling coefficient but increase with the magnetic damping coefficient. These results deepen the understanding of the mechanisms of acoustic attenuation in magnetic crystals and provide insights into the design of new-concept spin interconnects that operate based on acoustically driven magnon propagation.

*E-mail: jhu238@wisc.edu

## I. Introduction

Magnetic insulators are promising candidates for the interconnect application in spin wave (magnon) computing devices due to the dissipation-free magnon propagation [1–6]. Non-local transport of incoherent magnons has been demonstrated in low-damping ferrimagnetic insulator films such as yttrium iron garnet ($Y_3Fe_5O_{12}$, or YIG) [7–10] and spinel ferrite $MgAl_{0.5}Fe_{1.5}O_4$ [11] and the antiferromagnetic insulator film such as haematite ($\alpha$-$Fe_2O_3$) [12]. However, the intrinsic magnon propagation length in these materials, which is on the order of 10 μm, still falls short of the >100 μm goal for practical spin interconnect application [13]. A promising approach is to use a traveling acoustic wave (coherent phonons) to generate coherent magnons via the magnon-phonon interaction, thereby extending the magnon propagation length to the acoustic wave propagation length. This approach was computationally proposed based on advanced micromagnetic-elastodynamic simulations [14–18] and recently demonstrated by experiment, where a propagation distance of about 6 millimeters for gigahertz (GHz) coherent magnons was observed even in magnetic metal Ni with relatively large magnetic damping [19].

Despite these exciting developments, the current understanding of the acoustic attenuation in magnetic crystals, which determines the acoustic wave propagation length, is far from complete. In particular, a quantitative understanding of the dynamics of acoustic wave attenuation is still missing when magnons and phonons of similar frequencies and wavenumbers hybridize into magnon polarons [20]. In this case, the injected acoustic wave would be converted to a magnetoelastic wave (its quanta are magnon polarons) whose energy is converted back-and-forth between the spin and lattice subsystems with nearly 100% efficiency during propagation. The damping of such magnetoelastic wave, which is strongly determined by both the magnetic and elastic damping, is challenging to address by analytical calculation [21]. Although there are a few advanced computational models [14–18] which consider the generation of secondary acoustic waves from the acoustically excited spin waves and hence in principle permit modeling magnon polaron formation, the features of magnon polaron formation were not reported in those works. We suggest this is either because the frequency $f$ and wavenumber $k$ of the acoustic and spin waves do not match the condition of magnon polaron formation close enough, or the magnetic damping coefficient used in these computation studies is too large (which was also suggested in [18]).

In this article, we computationally investigate the effects of magnon polaron formation on the attenuation of a bulk acoustic wave in a magnetic insulator film, which has not yet been addressed in existing computational or experimental works. Time-domain simulations were performed based on an in-house Graphics Processing Unit (GPU)-accelerated dynamical phase-field model [17,22]. This model not only considers the two-way coupled dynamics of acoustic waves and spin waves in magnetic thin-film heterostructures like in the existing works [14–18], but also includes the previously omitted elastic stiffness damping for achieving a more accurate modeling of acoustic attenuation. In particular, we demonstrate the existence of a minimum magnon-phonon interaction time for the magnon polaron formation, which is strongly influenced by the magnetoelastic coupling and magnetic damping coefficients. Our results deepen the understanding of the mechanisms of acoustic attenuation in magnetic crystals and provide insights into the design of new-concept spin interconnects that operate based on acoustically driven magnon propagation. Moreover, the time-domain dynamics of the formation, transport, and attenuation of magnon polaron studied herein can be extended to other hybrid magnonic systems [23–26] that involve the hybridization between magnons and other quasiparticles such as photons [27,28] and qubits [29].

## II. Analytical Formulation of the Magnon Polaron Dispersion Relation

The formation of a magnon polaron requires (1) the magnons and phonons have identical or at least similar frequencies (*f*) and wavenumbers (*k*); (2) the values of *f* and *k* match close enough the values at the crossing points of the magnon and phonon dispersion curves; (3) both the magnetic and elastic damping need to be low enough; (4) the interaction time between magnons and phonons is long enough. To our knowledge, part (4) has previously not been explicitly stated. Due to these multiple requirements, there are only a few reports on the direct experimental observation of magnon polarons [30–35]. In fact, a majority of these works are limited to the observation of *k*=0 mode magnon polaron [30–33], which describe the coupled evolution of spatially uniform magnetization (*k*=0 mode magnon) and mechanical displacements (*k*=0 mode phonon). Although such *k*=0 mode magnon polaron can exist in systems with relatively large magnetic damping such as Ni [30] and $Fe_{81}Ga_{19}$ [31], the observation of *k* ≠0 mode magnon polaron, which is more relevant to the spin interconnect application, has thus far only been observed in epitaxial iron garnet thin films that has low magnetic damping [34,35]. For this reason, we consider (001) yttrium iron garnet ($Y_3Fe_5O_{12}$, or YIG) film as a representative material in this work. In this section, we analytically calculate the magnon polaron dispersion relation of the (001) YIG film, which will then be utilized to guide the dynamic phase-field simulations.

Let us consider the equilibrium magnetization $\mathbf{m}^0$ of the (001) YIG film is along +*x*, which is stabilized by a bias magnetic field $\mathbf{H}^{bias}$ applied along the same direction. For such an in-plane-magnetized (001) YIG film, only a transverse acoustic wave such as $\varepsilon_{xz}(z,t)$ and $\varepsilon_{yz}(z,t)$ can rotate the local magnetization due to the symmetry of the magnetoelastic anisotropy field [17,36]. Therefore, we consider a continuous bulk acoustic wave $\varepsilon_{xz}(z,t)$, which can be generated by a piezoelectric transducer [16,18,37], propagates into the YIG film from its bottom surface (*z*=0), and an adjacent gadolinium gallium garnet ($Gd_3Ga_5O_{12}$, or GGG) substrate works as the sink of the acoustic wave, as shown in Fig. 1a. The dispersion relation of the magnon polarons (*f*–*k*) formed through the hybridization of the transverse acoustic (TA) phonons and acoustically excited magnons can be analytically determined by linearizing the coupled equations of motions for the local magnetization **m** and local mechanical displacement **u**, given as,

$$\omega = \sqrt{\frac{\omega_m^2+\omega_{ph}^2}{2} \pm \sqrt{\left(\frac{\omega_m^2-\omega_{ph}^2}{2}\right)^2 + \frac{B_2^2\gamma^2 q^2}{\mu_0 M_s \rho}\left(H_x^{bias}+Dq^2+\frac{2K_1}{\mu_0 M_s}\right)}}, \quad (1)$$

where $\omega = 2\pi f$ is the angular frequency of the magnon polarons. $\omega_m$ and $\omega_{ph}$ are the angular frequencies of the magnons and phonons, respectively. Both the $\omega_m$ and $\omega_{ph}$ are functions of the angular wavenumber $q = 2\pi k$ and their expressions are given in Supplementary Material 1. $B_2$ is the magnetoelastic coupling coefficient, $\gamma$ is gyromagnetic ratio, $\mu_0$ is vacuum permeability, $M_s$ is saturation magnetization, $\rho$ is mass density, $D = \frac{2A_{ex}}{\mu_0 M_s}$ is exchange stiffness where $A_{ex}$ is the exchange coupling coefficient, and $K_1$ is magnetocrystalline anisotropy coefficient. Details of deriving Eq. (1) are also shown in Supplementary Material 1. The formation of magnon polarons induces two anticrossing in the dispersion relations of the TA phonons and exchange-coupling-dominated magnons, as shown in Fig. 1b. Figure 1c shows the dispersion relation of the magnon polaron at the low-*k* anticrossing at $\mathbf{H}^{bias}$ = 2605 Oe, with an exceptional point (where the dispersions of the TA phonon and magnon cross) of $(k_0, f_0)$ = (2.6 µm$^{-1}$, 10 GHz). The magnon-phonon coupling strength is defined as the wavenumber splitting [24] and denoted as $\Delta k$ in Fig. 1c. It is worth noting that that such analytically calculated dispersion relation of magnon polarons, like existing theories on this topic [21,38–41], are obtained based on two key assumptions: (i) both the magnetic damping and elastic damping are zero; (ii) the acoustic and spin waves both propagate in an infinitely long media without reflection. A nonzero magnetic/elastic damping will reduce the frequency/wavenumber gap at the anticrossing, leading to a smaller coupling strength. The reflection of the acoustic and spin waves at the heterostructure interface or film surface will make the value of coupling strength different from the analytical prediction.

### III. Dynamical Phase-field Model

Based on the heterostructure shown in Fig. 1a, we numerically model the influence of the magnon polaron formation on the attenuation of acoustic wave in the YIG film. The simulations were performed using an in-house GPU-accelerated dynamical phase-field model that considers the coupled dynamics of acoustic phonons, magnons, photons, and plasmons in magnetic thin-film heterostructures [22]. For the present problem, it is not necessary to consider the coupling to the dynamics of photons because the radiation magnetic field produced by the magnons is negligibly small (See Supplementary Material 2). It is also not necessary to consider the coupling to the dynamics of plasmons (free electron gas) [40] since the YIG is an electronic insulator.

In our phase-field model, the temporal evolution of the normalized local magnetization $\mathbf{m}$ in the YIG is governed by the Landau-Lifshitz-Gilbert (LLG) equation,

$$\frac{\partial \mathbf{m}}{\partial t} = -\frac{\gamma}{1+\alpha^2}\mathbf{m}\times\mathbf{H}^{\text{eff}} - \frac{\alpha\gamma}{1+\alpha^2}\mathbf{m}\times(\mathbf{m}\times\mathbf{H}^{\text{eff}}), \qquad (2)$$

where $\alpha$ is the effective magnetic damping coefficient. The total effective magnetic field $\mathbf{H}^{\text{eff}}$ is a sum of the magnetocrystalline anisotropy field $\mathbf{H}^{\text{anis}}$, the Heisenberg exchange coupling field $\mathbf{H}^{\text{exch}}$, the magnetic dipolar coupling field $\mathbf{H}^{\text{dip}}$, the bias magnetic field $\mathbf{H}^{\text{bias}}$ (fixed along +$x$ in this work), the magnetoelastic anisotropy field $\mathbf{H}^{\text{mel}}$. The mathematical expressions of $\mathbf{H}^{\text{anis}}$ and $\mathbf{H}^{\text{dip}}$ (both are a function of $\mathbf{m}$), $\mathbf{H}^{\text{exch}}$ (a function of $\nabla^2\mathbf{m}$), and the $\mathbf{H}^{\text{mel}}$ (a function of $\mathbf{m}$ and local strain $\boldsymbol{\varepsilon}$) are provided in our previous work [22].

The injected acoustic wave $\boldsymbol{\varepsilon}(z,t)$, which is considered to be spatially uniform in the $xy$ plane, excites the spin wave $\mathbf{m}(z,t)$ via the $\mathbf{H}^{\text{mel}}$. Therefore, the coupled transport of acoustic and spin waves occurs along the $z$ axis and can be modelled in a one-dimensional (1D) system [16,18,22]. The strain is calculated as $\varepsilon_{ij}=\frac{1}{2}\left(\frac{\partial u_i}{\partial j}+\frac{\partial u_j}{\partial i}\right)$ ($i, j = x, y, z$) and the evolution of the mechanical displacement $\mathbf{u}$ is described by the elastodynamics equation [42,43],

$$\rho\frac{\partial^2 \mathbf{u}}{\partial t^2} = \nabla\cdot\left(\boldsymbol{\sigma}+\beta\frac{\partial\boldsymbol{\sigma}}{\partial t}\right), \qquad (3)$$

where stress $\boldsymbol{\sigma}=\mathbf{c}(\boldsymbol{\varepsilon}-\boldsymbol{\varepsilon}^0)$; $\rho$, $\beta$ and $\mathbf{c}$ are the mass density, stiffness damping coefficient and elastic stiffness, respectively, for the (001)-oriented YIG or GGG. The $\boldsymbol{\varepsilon}^0$ is stress-free strain caused by the magnetization via magnetostriction, and $\varepsilon_{ii}^0=\frac{3}{2}\lambda_{100}^{\text{M}}\left(m_i^2-\frac{1}{3}\right)$, $\varepsilon_{ij}^0=\frac{3}{2}\lambda_{111}^{\text{M}}m_im_j$, with $i,j=x,y,z$, where $\lambda_{100}^{\text{M}}$ and $\lambda_{111}^{\text{M}}$ are the magnetostrictive coefficients of the YIG. For a 1D system whose physical quantities are uniform in the $xy$ plane, the full expansion of Eq. (3) is given as,

$$\rho \frac{\partial^2 u_x}{\partial t^2} = \left(1+\beta\frac{\partial}{\partial t}\right)\left[c_{44}\frac{\partial^2 u_x}{\partial z^2} + B_2\frac{\partial(m_x m_z)}{\partial z}\right], \tag{4a}$$

$$\rho \frac{\partial^2 u_y}{\partial t^2} = \left(1+\beta\frac{\partial}{\partial t}\right)\left[c_{44}\frac{\partial^2 u_y}{\partial z^2} + B_2\frac{\partial(m_y m_z)}{\partial z}\right], \tag{4b}$$

$$\rho \frac{\partial^2 u_z}{\partial t^2} = \left(1+\beta\frac{\partial}{\partial t}\right)\left[c_{11}\frac{\partial^2 u_z}{\partial z^2} + B_1\frac{\partial(m_z^2)}{\partial z}\right], \tag{4c}$$

where $B_1=-1.5\lambda_{100}^{M}(c_{11}^{M}-c_{12}^{M})$ and $B_2=-3\lambda_{111}^{M}c_{44}^{M}$ are the magnetoelastic coupling coefficients of the YIG. Equation 4(a-c) indicate that the precession of local magnetization can generate secondary acoustic waves via the terms which are related to $B_1$ and $B_2$ (namely, the magnetoelastic stress). The entire YIG/GGG heterostructure is discretized into a 1D system of computational cells along the $z$ axis, with a cell size $\Delta z=2$ nm (slightly larger than the unit cell size of YIG, ~1.2 nm [44]). The injection of the continuous bulk acoustic wave $\varepsilon_{xz}(z,t)$ is simulated by applying a time-varying mechanical displacement, $u_x(z=0, t)=u_{\max}\sin(2\pi f_{app}t)$, at the bottom surface ($z=0$) of the YIG film, where $f_{app}$ is the frequency of the injected acoustic wave. Central finite difference is used for calculating the spatial derivatives. All equations are solved simultaneously using the classical Runge-Kutta method for time-marching with a real-time step $\Delta t=20$ fs. The magnetic boundary condition $\partial \mathbf{m}/\partial z=0$ is applied on the two surfaces of the YIG film. When solving the Eq. (3), the continuity boundary conditions of mechanical displacement $\mathbf{u}$ and stress $\boldsymbol{\sigma}$ (see details in [17]) are applied at the YIG/GGG interface. The absorbing boundary condition, $\frac{\partial u_i}{\partial z}=-\frac{1}{v}\frac{\partial u_i}{\partial t}$ ($i=x,y,z$), is applied at the top surface of the GGG substrate to make it a perfect sink for acoustic waves. Here $v$ is the transverse sound velocity for $u_x$ and $u_y$ and the longitudinal sound velocity for $u_z$. All solvers are GPU-accelerated to achieve high-throughput simulations in a computational system of the order of $10^5$ cells and millions of numerical time steps.

The materials parameters are summarized as follows. For the (001) GGG [45], the elastic stiffness coefficients $c_{11} = 285.7$ GPa, $c_{12} = 114.9$ GPa, $c_{44} = 90.2$ GPa and $\rho = 7085$ kg m$^{-3}$. For the (001) YIG [18,41,46], $c_{11} = 269$ GPa, $c_{12} = 107.7$ GPa, $c_{44} = 76.4$ GPa and $\rho = 5170$ kg m$^{-3}$; $\gamma = 0.22$ rad MHz A$^{-1}$ m; $M_s = 0.14$ MA m$^{-1}$; $A_{ex}=3.26$ pJ m$^{-1}$; $K_1 = 602$ J m$^{-3}$; $B_1 = 0.3$ MJ m$^{-3}$ and $B_2 = 0.55$ MJ m$^{-3}$. The stiffness damping coefficient of the (001) YIG is assumed to be same as that of the GGG, i.e., $\beta = 3\times10^{-15}$ s. The value of $\beta$ is obtained by fitting the experimentally determined characteristic decay length of a GHz transverse acoustic wave (~ 2 mm) in (001) GGG [32] (see details in Supplementary Material 3).

## IV. Results and Discussion

Guided by analytical calculation (Fig. 1c), a continuous acoustic wave $\varepsilon_{xz}$ with an amplitude of $10^{-5}$ ($u_{\max}=1.224$ pm) and a frequency $f_{app}$ of 10 GHz is injected into the (001) YIG film. To determine the existence of magnon polarons, we numerically extract the magnon-phonon coupling strength – the magnitude of wavenumber splitting or shift $|\Delta k|$ – from the simulated spatial profile of the acoustic wave. A nonzero $\Delta k$ indicates the presence of magnon polaron. A 200-µm-thick YIG film is considered as the main example, which allows us to obtain a 200-µm-long profile for both the spin wave and acoustic wave without wave components reflected from the YIG/GGG interface. As a result, the numerically extracted $\Delta k$ from such spatial profiles can be utilized for comparison to the analytical solution in Fig. 1c which was also obtained by assuming no wave reflection. Figure 2a compares the spatial profiles of the acoustically excited spin wave $m_y(z)$ and the injected acoustic wave $\varepsilon_{xz}(z)$ in the 200-µm-thick YIG film at the moment ($t$=52 ns) when the acoustic wave just arrives at the YIG/GGG interface. As seen, the amplitude of acoustic wave reduces to almost zero at the nodes where the spin wave amplitude is still significant if not at a local maximum, indicating an elastic-to-magnetic energy conversion of nearly 100% efficiency. From their corresponding wavenumber spectra in Fig. 2b, the wavenumber splitting from the 2.6 µm$^{-1}$ (the $k$ value of the injected acoustic wave) to the two distinct values of 2.57 µm$^{-1}$ and 2.64 µm$^{-1}$ is clear, indicating the formation of magnon polarons. We further vary the $f_{app}$ from 9.8 GHz to 10.2 GHz in the simulations. As shown in Fig. 2c, the numerically extracted coupling strength $|\Delta k|$ agrees well with the analytical prediction, demonstrating the high numerical accuracy of our in-house dynamical phase-field model. In the case of thinner YIG films (e.g., 10 µm), although the magnon polaron can still form after the magnons and phonons interact for a sufficiently long time, the numerically extracted $|\Delta k|$ is different from the analytical prediction due to the wave reflection (see details in Supplementary Material 4). However, when the YIG film thickness is comparable to or smaller than the single wavelength of the acoustic wave, which is 384 nm for a 10-GHz $\varepsilon_{xz}$ in (001) YIG, $k = 0$ mode magnon polaron will appear instead of the $k \neq 0$ mode magnon polaron.

To evaluate the influence of magnon polaron formation on the acoustic attenuation, we define the attenuation ratio $h$ of the injected acoustic wave as $h=1-\int_0^d \varepsilon_{xz}(z)^2 dz / \int_0^d \varepsilon_{xz}^{uni}(z)^2 dz$, where the integration evaluates the elastic energy of the entire acoustic wave packet; $\varepsilon_{xz}^{uni}(z)$ refers to the acoustic wave profile simulated by omitting the magnetoelastic stress (see Supplementary Material 5); $d$=200 µm is the YIG film thickness. Figure 2d shows the variation of $h$ with the frequency of the injected acoustic wave $f_{app}$, which shows a similar trend to that of $|\Delta k|$ in Fig. 2c. When the frequency $f_{app}=f_0$=10 GHz, $h$ reaches its peak value of ~90%.

Now we discuss the influence of the magnetic damping coefficient on the magnon polaron formation and the acoustic wave attenuation. Experimentally, it has been shown that the magnetic damping coefficient of single-crystal YIG can be enhanced to close to 0.1 through Bi-doping without significant changes in the other materials parameters such as the saturation magnetization [47]. Figure 3a shows the spatial profiles of the injected acoustic wave ($f_{app}$=10 GHz) in the 200-µm-thick YIG films with the magnetic damping coefficient $\alpha = 0.001$, 0.002 and 0.1 at $t$=52 ns, with comparison to the profile in the bottom panel of Fig. 2a where $a$=8×10$^{-5}$. Here, we focus on the first 50 µm of the acoustic wave packet ($z$=0-50 µm) where the phonon and magnon had interacted for long enough time (up to 52 ns). As shown by their corresponding wavenumber spectra in Fig. 3b, the formation of magnon polaron (a nonzero $|\Delta k|$) is seen only in the case of $\alpha = 0.001$. This is because a larger magnetic damping would lead to a larger amount of energy dissipation and ultimately suppress the magnon polaron formation.

Figure 3c presents the attenuation ratio $h$ of the acoustic waves $\varepsilon_{xz}(z,t)$ as a function of the $\alpha$ where the $f_{app}$ varies from 9.8 GHz to 10 GHz. Given that the magnon-phonon coupling strength $|\Delta k|$ increases monotonically to its maximum when $f_{app}$ increases from 9.8 GHz to 10 GHz (see Fig. 2c), two conclusions can be made. First, the larger the $|\Delta k|$ is, the higher the acoustic attenuation. This is because stronger coupling strength enables the conversion of more elastic energy into the magnetic energy. As shown in Fig. 3c, the curve with $f_{app}$=10 GHz is above all the other curves with lower $f_{app}$. Second, for a given value of $f_{app}$, the highest acoustic attenuation appears at an intermediate value of *a*. For example, *h* reaches its peak of 96.5% at $a=0.001$ for $f_{app} =10$ GHz. This is due to the following two competing effects. On one hand, a larger *a* leads to a smaller $|\Delta k|$ (shown in Fig. 3b) and hence lower acoustic attenuation. On the other hand, it yields a more significant magnetic energy dissipation, which would lead to higher acoustic attenuation because fewer amount of magnetic energy is converted back to the elastic energy.

Our time-domain simulations also allow for evaluating the minimum time required for the magnon polaron formation, denoted as $t_{min}$, which is important to both the fundamental understanding of the acoustically driven magnon propagation and the design of the novel spin interconnect based on this phenomenon. However, such $t_{min}$ is hardly possible to determine with classical analytical theories and has not yet been investigated by computation or experiment. Here we evaluate the $t_{min}$ based on the moment the initially zero $|\Delta k|$ starts to increase significantly. Figure 4a shows the spatial profiles of the $\varepsilon_{xz}(z,t)$ at $t= 7.8$ ns, 13 ns, and 26 ns in the 200-µm-thick YIG film. As shown by their wavenumber spectra in Fig. 4b, the $|\Delta k|$ is still zero at $t= 7.8$ ns with a single peak wavenumber at 2.6 µm$^{-1}$. As the time increases to 13 ns and 26 ns, the $|\Delta k|$ is nonzero, indicating the formation of magnon polarons. By tracing the peaks in the wavenumber spectra, we plot the temporal evolution of $|\Delta k|$ in Fig. 4c, where it is found that the $|\Delta k|$ becomes nonzero at $t \sim 9$ ns, indicating the formation of magnon polarons. $|\Delta k|$ then increases monotonically and saturates at $t \sim 13$ ns. Note that the saturation value of $|\Delta k|$ (~0.033 µm$^{-1}$) is smaller than its theoretical upper bound of 0.04 µm$^{-1}$ (c.f., Fig. 1c). This is due to either the nonzero (albeit small) magnetic and elastic damping coefficients used in the dynamical phase-field simulations and/or the numerical noise from the discrete Fourier transformation used for obtaining the wavenumber spectra. Moreover, since the onset of nonzero $|\Delta k|$ should be associated with the appearance of the nodes in an acoustic wave packet (where spin wave amplitude is significant or maximized, as discussed in Fig. 2a), we also plot the spatiotemporal profile of the $u_x(z,t)$ during the first 52 ns in the 200-µm-thick YIG film. As shown in Fig. 4d, the node starts to appear at $t=9.6$ ns, which is almost the same as the moment when the $|\Delta k|$ starts to increase significantly in Fig. 4c. More detailed analyses indicate that such a threshold duration $t_{min}$ decreases as the magnetoelastic coupling coefficients increase (see Supplementary Material 6), and increases when the *a* gets larger (see Supplementary Material 7). However, if the *a* is too large, the formation of magnon polarons would not be possible since the enhanced energy dissipation would diminish the coherent energy transfer between the spin and lattice subsystems [24]. Furthermore, our control simulations show that the $t_{min}$ is barely influenced by the wave reflection at the interfaces. For example, in the case of the 10-µm-thick YIG film, our simulations reveal a similar $t_{min}$ of ~9.2 ns (see Supplementary Material 4) despite the strong interference between the incident and reflected spin waves.

The knowledge of $t_{min}$ allows us to design a fast electrical switch that permits a spatially precise, ns-scale control of the acoustic/spin wave profile by applying a local dynamical magnetic field $\mathbf{H}^{Dyn}$ via a patterned microstrip. As illustrated in Fig. 5a, such an envisaged scheme can incorporate a dynamical tuning capability into the earlier-mentioned concept of spin interconnect that operates based on acoustically driven spin wave propagation. The principle of such dynamical tuning is clear: varying the magnetic field can shift the frequency of the exceptional point $f_0$ (see Fig. 1b) and therefore turn the system on/off the resonance. For quasi-static magnetic-field control [48], the knowledge of $t_{min}$ is not critical. However, the acoustic/spin wave profile can be very sensitive to the duration of $\mathbf{H}^{Dyn}$ when the latter is at the ns-scale, particularly in the vicinity of the $t_{min}$.

To provide a proof-of-principle demonstration of such dynamical tuning of both the acoustic and spin wave, we simulate the time-domain evolution of acoustic wave profiles $\varepsilon_{xz}(z,t)$ across a (001) YIG(20 µm)/GGG(substrate) and the acoustically excited spin wave in the YIG under pulsed $\mathbf{H}^{Dyn}$. Likewise, a 10-GHz continuous acoustic wave was injected and the magnetic damping coefficient $a$ is $8\times10^{-5}$. The bias magnetic field $\mathbf{H}^{bias}$=2000 Oe. Both the $\mathbf{H}^{Dyn}$ and the $\mathbf{H}^{bias}$ are along the $+x$ direction. Figure 5b-d shows the temporal evolution of the local strain $\varepsilon_{xz}(t)$ and the local magnetization $m_y(t)$ at the YIG/GGG interface under pulsed $\mathbf{H}^{Dyn}$ of three different durations. In all three cases, when $\mathbf{H}^{Dyn}$ reaches its maximum, the YIG film would be on magnon-phonon resonance (i.e., $f_0=f_{app}$=10 GHz). As shown in Fig. 5b, the influence of $\mathbf{H}^{Dyn}$ on the acoustic wave amplitude is not appreciable when the maximum value of $\mathbf{H}^{Dyn}$ is only held transiently. The amplitude of the acoustically excited spin wave is relatively weak since the system is only on resonance transiently. When $\mathbf{H}^{Dyn}$ is held for longer time at its maximum value, the acoustic attenuation is more significant, as shown in Fig. 5c, yet the spin wave has a larger amplitude since the system is on resonance for a longer time. When the duration of the $\mathbf{H}^{Dyn}$ is long enough to enable the formation of magnon polarons, significant acoustic wave attenuation occurs over the entire 'ON' phase of the pulsed $\mathbf{H}^{Dyn}$ with clear back-and-forth oscillation in amplitude (see Fig. 5d), which is a signature feature of strong coherent magnon-phonon coupling in the time-domain known as the coherent beating oscillation [34]. The acoustically excited spin wave shows a largely complementary oscillatory feature in the time domain and has the largest amplitude due to the high energy conversion efficiency from the high magnon-phonon coupling strength.

**IV. Conclusions**

In conclusion, we have performed dynamical phase-field simulations to model the attenuation of a bulk acoustic wave in a magnetic insulator film when this injected acoustic wave is converted to a magnetoelastic wave (the formation of magnon polarons). Our simulation results show that the acoustic attenuation in magnetic insulator films is larger under a stronger magnon-phonon coupling strength which is indicated by the magnitude of wavenumber splitting/shift $|\Delta k|$, where a nonzero $|\Delta k|$ indicates the formation of magnon polaron. Yet, perhaps somewhat counterintuitively, the results show that the attenuation is the strongest when the effective magnetic damping coefficient $a$ takes an intermediate value. Furthermore, the results also demonstrate the existence of a minimum interaction time between magnons and phonons ($t_{min}$) required for the magnon polaron formation, and reveal how the $t_{min}$ is influenced by key materials parameters such as the magnetoelastic coupling and magnetic damping coefficients. These simulation results can be utilized to guide the materials and heterostructure design to achieve a spatially precise, ns-scale magnetic-field control of acoustic/spin wave profiles for spin interconnect applications. The conceptual understanding of the minimum interaction time $t_{min}$ can be extended to the hybridization between magnons and other quasiparticles such as the magnon-photon [27,28] and magnon-qubit [29] coupling.

Moreover, although a bulk transverse acoustic wave is considered in this work, the theoretical and numerical analyses can be extended to the interaction between the bulk longitudinal or surface acoustic waves and spin waves and to more complex magnetic heterostructures (e.g., superlattices, nanostructure arrays) [24,25,36,49,50]. Finally, beyond the magnetic systems, we note that the enhanced acoustic attenuation due to the magnon-phonon hybridization is analogous to the enhanced acoustic attenuation in ferroelectrics near the Curie temperature, which results from the strong coupling between the acoustic phonons and softened optical soft mode phonons [51–53].

**Acknowledgements**

J.-M.H. acknowledges support from the National Science Foundation (NSF) award CBET-2006028 and the Accelerator Program from the Wisconsin Alumni Research Foundation. The simulations were performed using Bridges at the Pittsburgh Supercomputing Center through allocation TG-DMR180076, which is part of the Extreme Science and Engineering Discovery Environment and supported by NSF grant ACI-1548562.

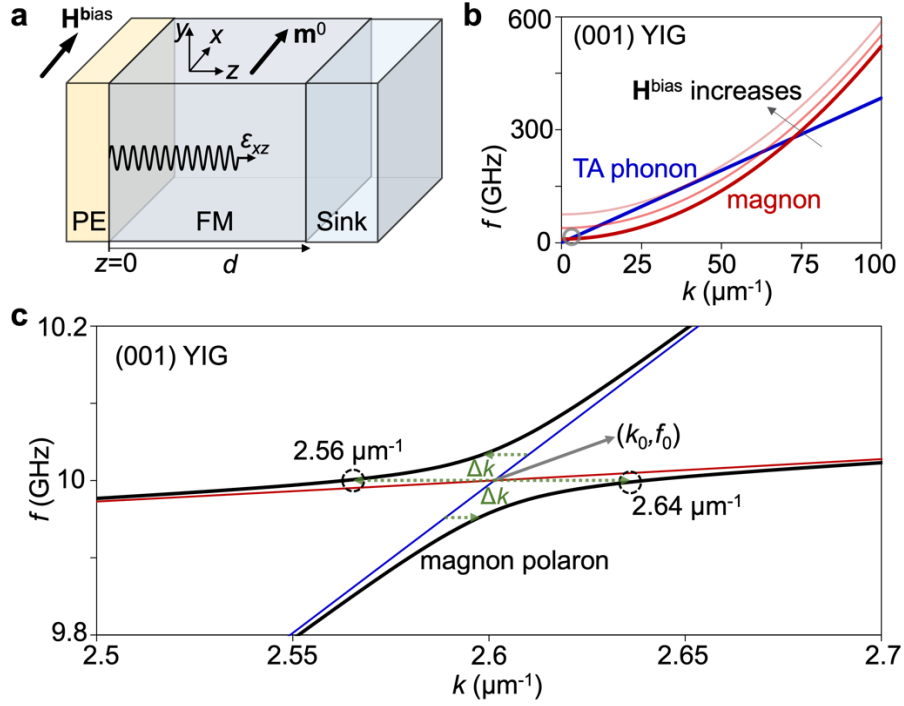

**Figure 1**. **a**, Schematic, not to scale, showing a piezoelectric (PE)/ferromagnet (FM)/substrate heterostructure, where the substrate is a sink for the continuous transverse acoustic wave $\varepsilon_{xz}$. A bias magnetic field $\mathbf{H}^{bias}$ is applied along $+x$ to stabilize a spatially uniform initial magnetization $\mathbf{m}^0$ along the same direction. **b**, Analytically calculated dispersion relations of transverse acoustic (TA) phonon (blue) and exchange-coupling-dominated magnons (red) in a (001) YIG film under $\mathbf{H}^{bias}=$ 2605 Oe, 13025 Oe, and 26050 Oe, where $\mathbf{k}\|z$. **c**, Magnon polarons at the low-wavenumber ($k$) anticrossing between the TA phonon and the magnon branches under $\mathbf{H}^{bias}=$ 2605 Oe. The formation of magnon polarons shifts the wavenumber of the injected acoustic wave by an amount indicated by $\Delta k$. $|\Delta k|$ is the largest at the exceptional point $(k_0, f_0) = (2.6\ \mu m^{-1}, 10\ GHz)$.

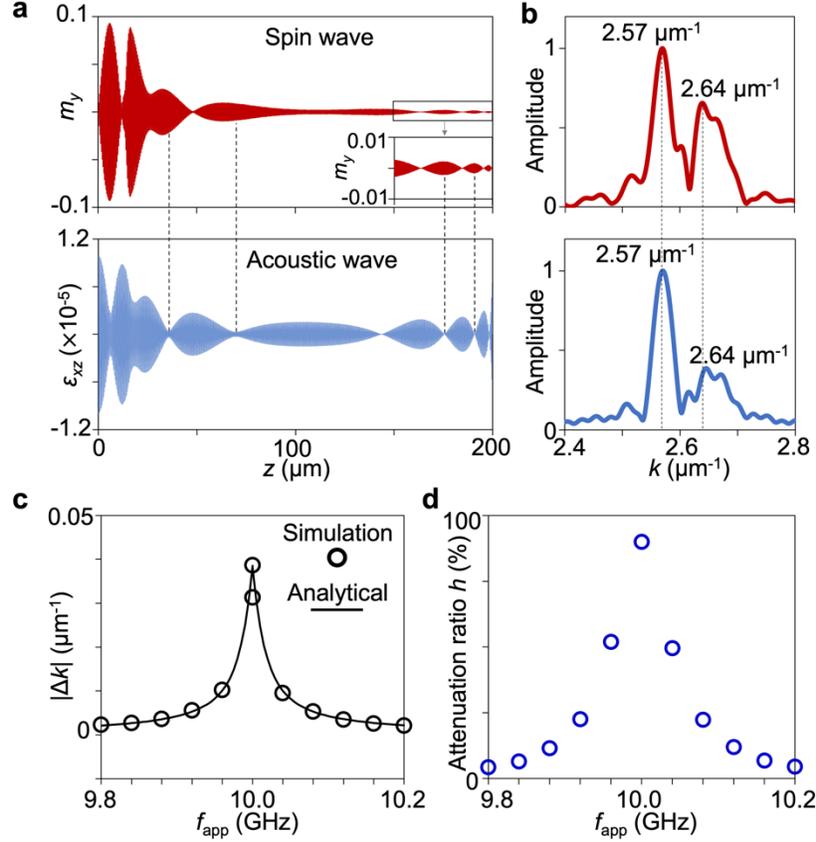

**Figure 2**. **a**, Spatial profiles of the spin wave $m_y(z)$ and the acoustic wave $\varepsilon_{xz}(z)$ in a 200-µm-thick (001) YIG film at $t = 52$ ns. $t=0$ ns is the moment that the acoustic wave propagates into the YIG film from its bottom surface ($z=0$). The inset shows the enlarged profile of the $m_y(z)$ in the last 50 µm. The dashed lines connect the local minima in the acoustic wave packet to the locations in the spin wave packet. **b**, Wavenumber spectra of the spin wave (top panel) and the acoustic wave (bottom panel) obtained by performing discrete Fourier transforms of the $m_y(z)$ and the $\varepsilon_{xz}(z)$ profiles shown in (**a**). **c**, Analytically calculated (solid lines) and numerically simulated (circles) wavenumber splitting/shift $|\Delta k|$ (representing the coupling strength) of the magnon polaron as a function of the frequency of the injected acoustic wave ($f_{app}$). **d**, Simulated acoustic wave attenuation ratio $h$ (see definition in the main text) as a function of the $f_{app}$.

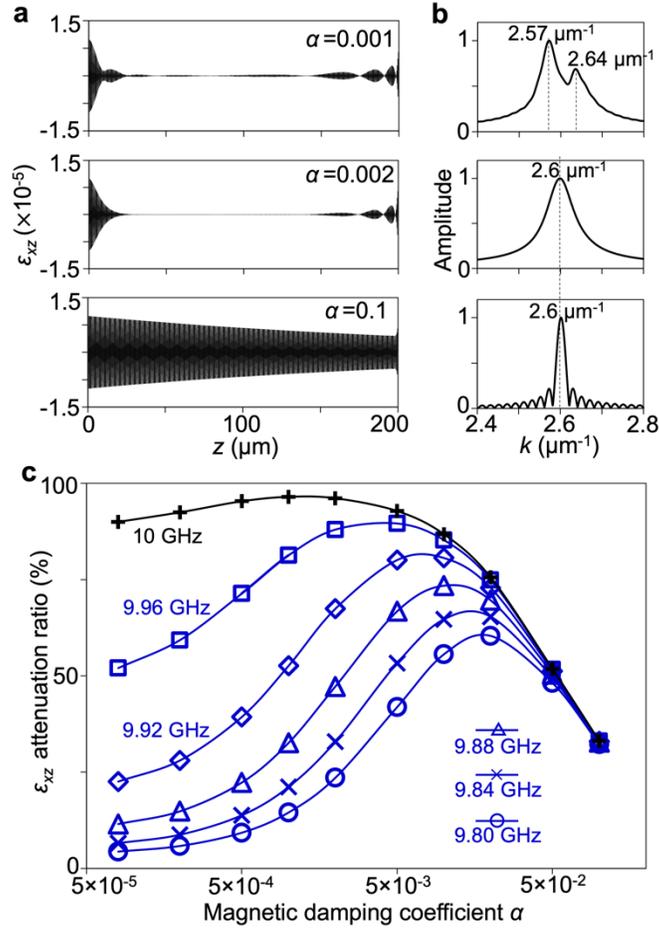

**Figure 3**. **a**, Spatial profiles of the acoustic wave $\varepsilon_{xz}(z)$ in a 200-µm-thick (001) YIG film at $t = 52$ ns with effective magnetic damping coefficient $\alpha = 0.001$, 0.002, and 0.1. **b**, Their wavenumber spectra obtained by performing discrete Fourier transforms of the first 50 µm of the $\varepsilon_{xz}(z)$ profile in (**a**). **c**, The acoustic wave attenuation ratio $h$ as a function of the $\alpha$, under different frequencies of the injected acoustic wave ($f_{\text{app}}$).

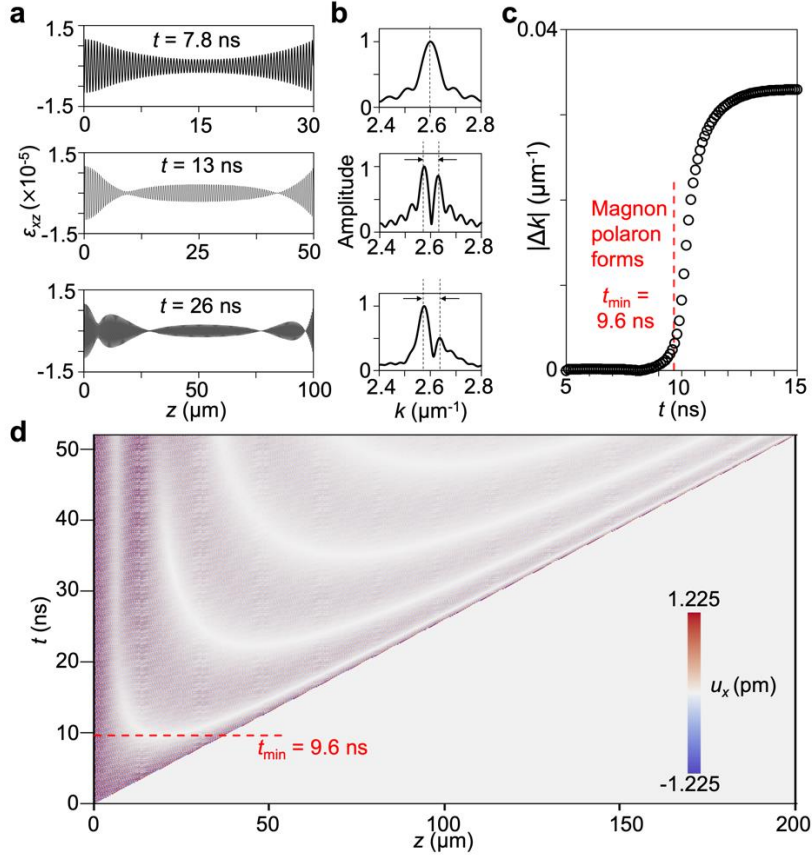

**Figure 4**. **a**, Spatial profiles of the $\varepsilon_{xz}(z,t)$ wave packet in a 200-µm-thick (001) YIG film at $t = 7.8$ ns, $t = 13$ ns, and $t = 26$ ns, and **b**, their wavenumber spectra. $t=0$ is the moment that the acoustic wave propagates into the YIG film from its bottom surface ($z = 0$). **c**, Temporal evolution of the amplitude of the negative wavenumber shift $|\Delta k| = |k-k_0|$ in **b**, where $k_0 = 2.6$ µm$^{-1}$ and $k<k_0$. **d**, Spatiotemporal profile of the mechanical displacement $u_x(z,t)$ in a 200-µm-thick (001) YIG film within $t = 0$-52 ns. The minimum time required for the magnon polaron formation $t_{min}$ is indicated.

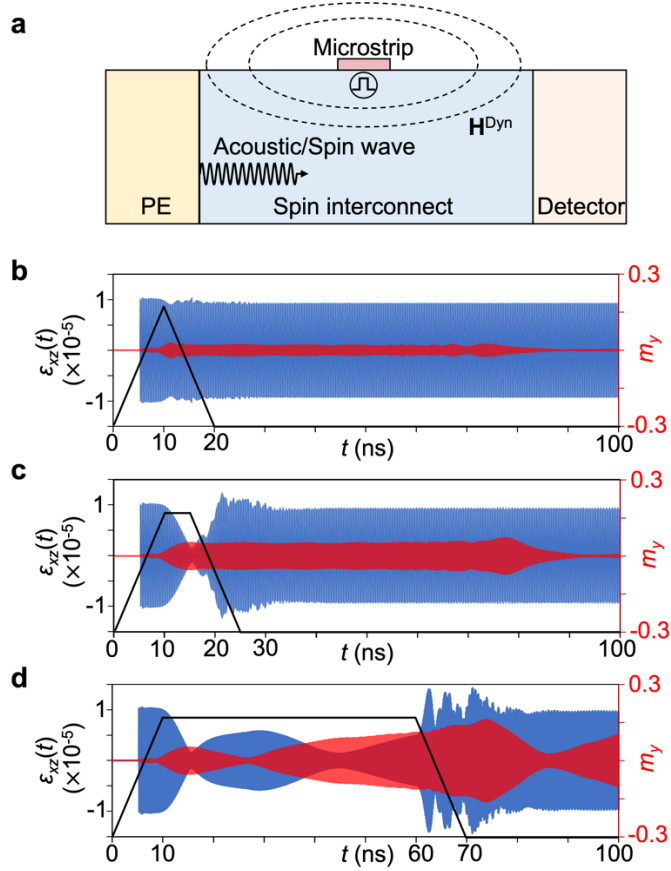

**Figure 5**. **a**, Schematic of an envisaged electrical switch for achieving ns-scale control of the injected acoustic wave and the acoustically excited spin wave for spin interconnect application. A piezoelectric (PE) transducer can be used for injecting acoustic wave. Pulsed dynamical magnetic field $\mathbf{H}^{Dyn}$ produced by the microstrip is used to shift the frequency of the exceptional point $f_0$ to turn the system on/off the magnon-phonon resonance. The detector can be a heavy metal that can detect the spin wave signal based on spin pumping and spin-charge conversion. **b–d**, Temporal profiles of the local strain $\varepsilon_{xz}(t)$ (blue) and magnetization $m_y(t)$ (red) at the YIG/GGG interface of a (001) YIG (20 μm)/GGG heterostructure, after injecting a 10-GHz continuous acoustic wave $\varepsilon_{xz}(t)$. The $\mathbf{H}^{Dyn}$ is held at its peak value of 605 Oe for (**b**) 0 ns, (**c**) 5 ns, and (**d**) 50 ns, as indicated by its temporal profiles shown by the black lines.